\documentstyle[prl,aps,twocolumn,epsfig]{revtex}
\draft

\begin{document}
\title{Exact versus mean-field description of the Bose-Einstein condensate:
a model study}

\author{Magdalena A. Za{\l}uska-Kotur, Mariusz Gajda, Arkadiusz Or{\l}owski,
and Jan Mostowski}
\address{Instytut Fizyki PAN \& College of Science,\\
Aleja Lotnik{\'o}w 32/46, 02-668 Warszawa, Poland}
\maketitle

\begin{abstract}
We study a system of trapped bosonic particles interacting by model harmonic forces.
Our model allows for detailed examination of the notion of an order
parameter (a condensate wave function). By decomposing a single particle
density matrix into coherent eigenmodes we study an effect of interaction on
the condensate. We show that sufficiently strong interactions cause that the
condensate disappears even if the whole system is in its lowest energy state.
In the second part of our paper we discuss the validity of the Bogoliubov
approximation by comparing its predictions with results inferred from the exactly soluble model.
In particular we
examine an energy spectrum, occupation, and fluctuations of the condensate. We
conclude that Bogoliubov approach gives quite accurate description of the
system in the limit of weak interactions.
\end{abstract}

\pacs{PACS number(s): 03.75.Fi, 05.30.Fk}

\section{Introduction}

Recent advances in the trapping techniques have renewed interest
in various aspects of many body theory. In fact a cloud of weakly
interacting trapped atoms is an ideal system for which various
aspects of many body theory can be tested and verified. The ideal
bosonic gas undergoes the Bose-Einstein condensation if the
phase-space density exceeds one. This phenomenon manifests itself
by the macroscopic occupation of the single particle ground state.
In the case of an interacting system the condensate wave function
can be defined by the spectral decomposition of the one-body
density matrix.  This decomposition is closely related to the
off-diagonal long range order \cite{Onsager} or the existence of
the order parameter, i.e., the `classical' field with given
amplitude and phase commonly used in the theory of superfluidity
\cite{Dalfovo,Griffin}. Realization of decomposition procedure is
practically impossible because it requires a full solution of the
many-body problem. Mean-field approaches are commonly used
instead. The basic idea for a mean-field description of the
dilute, weakly interacting Bose gas below transition temperature
was introduced by Bogoliubov \cite{Bogolubov}. Most of the results
for the interacting Bose-Einstein condensate are obtained within
the Bogoliubov theory which in many cases provides a reliable
quantitative description of the quantum Bose gas.  Indeed, the low
energy excitation spectrum of the trapped condensate well below
transition temperature \cite{Jin} agrees remarkably well with
predictions based on the Bogoliubov theory \cite{Edwards}. On the
other hand the mean-field Bogoliubov approach fails to reproduce
excitation spectrum at higher temperatures -- close to the
transition point \cite{Dodd}. Therefore, the question about limits
of validity of the Bogoliubov method is of great importance. One
possible way to test the quality of this approximation is to go
beyond the mean field theory. Another possibility to assess the
usefulness of Bogoliubov's theory is to study the exactly soluble
models and to compare their predictions with those based on the
approximate method. The models provide not only a unique soluble
many body problem but also allow to verify various approximations.
This will shed some light on the validity and exactness of the
Bogoliubov method, widely used in many body physics \cite{Fetter}.

There are only few exactly soluble models of quantum systems where the
interactions between atoms is chosen in the form allowing for the exact
analytic solution. These are: (i) the one-dimensional model of impenetrable
bosons introduced by Girardeau \cite{Girardeau}, (ii) its
contact potential version formulated by Lieb \cite{Lieb}; (iii) the model of
particles interacting by harmonic forces
\cite{Calogero,Bialynicki,Lemmens}. Although in the first two cases
the formal solution is given but in practice the problem is still quite
complicated and quantitative calculations can be done for a very small number
of particles
only \cite{Rzazewski}. The latter case seems to be much simpler because, as it
has been shown in \cite{Magda}, it can be reduced to the problem of
noninteracting particles in a harmonic trap. Therefore in the following we
are going to examine, within this exactly soluble model, various concepts and
methods related to the interacting Bose-Einstein condensate.

This paper is organized as follows. In Sec.\ II we present exact
results regarding properties of harmonically interacting bosons
trapped within harmonic potential. The results are obtained within
a soluble model that was developed in \cite{Magda}. In the second
part of Sec.\ II we find the analytic expression for the order
parameter and study the effect of quantum depletion of the
condensate as well as quantum fluctuations at zero temperature. In
the third part we analyze the thermal properties of the system. In
Sec.\ III we apply the Bogoliubov method to our model. Within this
approximation we first determine a condensate wave function and an
excitation spectrum.  Using the Bogoliubov spectrum we calculate
occupation of the condensate and its fluctuation at finite
temperatures. We compare these mean-field results to the results obtained within the exact model.  We finish in Sec.\ IV with some concluding remarks.

\section{Exact results}

\subsection{Ground state and excitation spectrum}
In our previous paper \cite{Magda} we have shown the algebraic
method of diagonalization of the Hamiltonian describing a system
of many particles interacting via harmonic forces. The system
under consideration consists of many particles confined by an
external harmonic potential interacting by harmonic forces, i.e.,
two body interaction potential has the form:
\begin{equation}
V({\bf x}_i - {\bf x}_j) = \frac{\sigma}{2}\Omega^2 ({\bf x}_i - {\bf
x}_j)^2,
\end{equation}
where $\Omega$ defines the interaction strength and $\sigma = +1$ signifies the
attractive interaction of particles placed at positions ${\bf x}_i$ and ${\bf
x}_j$ whereas $\sigma =-1$ -- corresponds to repulsive interactions.
The total Hamiltonian of the $N$-particle system has therefore the following
form:
\begin{equation}
H = \sum_{i=1}^N \frac{1}{2}({\bf p}_i^2 + {\bf x}_i^2)
+ \sum_{i<j} V({\bf x}_i - {\bf x}_j).
\label{a0}
\end{equation}

Let us first recall the exact results of \cite{Magda}. For the sake of
simplicity we denote the set of all particle positions vectors
by ${\bf X}_N = ({\bf x}_1, \ldots, {\bf x}_N)$.  The Hamiltonian can be easily
diagonalized if one introduces collective variables:
\begin{equation}
{\bf X}^c_N = {\cal Q_N}\, {\bf X}_N,
\end{equation}
where ${\bf X}^c_N = ({\bf x}^c_1, \ldots, {\bf x}^c_N)$
and the matrix ${\cal Q}_N = \{ q^N_{ij} \}$ is orthogonal.
One of these collective variables namely the center of mass of $N$-particle
system plays a particularly important role:
\begin{equation}
{\bf x}^c_N = \frac{1}{\sqrt{N}} \sum_{i=1}^N {\bf x}_i.
\end{equation}
The choice of $N-1$ remaining collective variables ${\bf X}^{c}_{N-1} =
({\bf x}^c_1, \ldots, {\bf x}^c_{N-1})$ is not unique
but this does not lead to any physical implications. In particular:
\begin{equation}
({\bf X}^c_{N-1})^2=\sum_{i=1}^{N-1} ({\bf x}^c_i)^2 = \sum_{i=1}^N {\bf x}^2_i - ({\bf x}^c_N)^2.
\end{equation}
In the following we are going to use a similar notation for description of a
subsystem of $s$-particles, $s=1, \ldots,N$.

The above-defined transformation brings the Hamiltonian to the
diagonal form and its eigenenergies can be easily found. While
determining a spectrum, however, one must take into account the
proper symmetry of a total wave function. In the case of bosonic
particles ($N > 2$) the allowed energies are:
\begin{equation}
\label{spec}
E=\left( \frac{3}{2} +m \right) + \left(\frac{3}{2} (N-1)+n \right)\omega,
\end{equation}
where $m=0,1,2\ldots$, $n=0,2,3\ldots$ and $\omega = \sqrt{1+
\sigma N\Omega^2}$. The first term describes excitations of the
center of mass, i.e., $d$-dimensional harmonic oscillator of
frequency equal to one. The second term in the Eq.\ (\ref{spec})
corresponds to excitations of $N-1$ relative degrees of freedom.
The frequency $\omega$ characterizes some effective potential felt
by an individual quasi-particle because it results from a combined
effect of all particles of our system. Let us observe that $\omega
= 1$ corresponds to the noninteracting case, the repulsive
interactions give $0< \omega < 1$ while  attractive forces lead to
$\omega > 1$. Moreover, very small values of $\omega \approx 0$
signify very strong repulsion which almost destabilizes the whole
system. It is very convenient to parameterize $\omega$ by an
exponent $\kappa$ defined in the following way:
\begin{equation}
\omega = N^{\kappa}.
\end{equation}
This exponent can be related to the actual strength of the interaction. In
fact, for  weakly interacting gas ($\omega \approx 1$) we  obtain very small
values of this parameter: $\kappa \approx 0$, while for strong interactions
($\omega \approx 0$ -- repulsion,  $\omega \gg 1$ -- attraction) we have
$|\kappa| \gg 1$.  Moreover, $\kappa$ is positive in the case of attraction
while it is negative for repulsion. Let us add at this point that in
realistic situations of  short-range interparticle interactions, large
Bose-Einstein condensates can exist only for  repulsive forces. In the
case of attraction the size of the trapped condensate is limited to about
1500 atoms \cite{Dalfovo}. In our oscillatory model the forces between
particles are negligible at small distances, therefore the model leads to the
condensation (in the thermodynamic limit) in both attractive and repulsive
case.

The ground state of the system is the following:
\begin{equation}
\label{wfun}
\Psi({\bf X}_N)=\Phi_0(\sqrt{\omega} {\bf X}^c_{N-1}) \Phi_0({\bf x}^c_N),
\end{equation}
where $({\bf X}^c_{N-1},{\bf x}^c_N) = {\cal Q}_N {\bf X}_N$ and
the function $\Phi_0(\sqrt{\omega} {\bf X}^c_{N-1})$ corresponds to the ground
state of a system of $N-1$ independent quasi-particles (in $d$ spatial
dimensions)
interacting with an external potential of the harmonic oscillator of frequency
$\omega$:
\begin{equation}
\label{phirel}
\Phi_0(\sqrt{\omega} {\bf X}^c_{N-1}) =
\left( \frac{\omega}{\pi} \right)^{d(N-1)/4}
{\rm exp}\left[-\omega ({\bf X}^c_{N-1})^2/2\right],
\end{equation}
and $\Phi_0({\bf x}^c_N)$ is the ground state of the single particle (center of
mass) trapped into harmonic potential:
\begin{equation}
\label{phicm}
\Phi_0({\bf x}^c_N) =
\left( \frac{1}{\pi} \right)^{d/4}
{\rm exp}\left[-({\bf x}^c_N)^2/2\right].
\end{equation}
Construction of excited eigenstates is difficult because it is not
easy to impose the desired symmetry on the wave function. Such a
procedure was describe in details in \cite{Magda}.

\subsection{Order parameter and quantum depletion}
If the energy of the system (or equivalently the temperature) is
sufficiently small we expect that the system forms a Bose-Einstein
condensate.  The BEC of the ideal gas manifests itself by a
macroscopic occupation of the single particle ground state. In the
case of interacting system it is not obvious what is this
particular state which is `macroscopically occupied'. The
identification of the macroscopically occupied quantum state is
equivalent to the definition of the order parameter -- the single
particle wave function which is inherently related to the Bose
condensation. The condensate subsystem can be then quite
accurately described by the $N_0$-fold product of the order
parameter, where $N_0 \simeq {\cal O}(N)$ is the occupation of the
condensate. In the conventional approaches, for example in the
Bogoliubov method, it is simply assumed that a mean value of the
boson field operator is different than zero and this mean value is
associated with the macroscopically occupied state. Then,
consistently with the above assumption, the Bogoliubov equations
give in fact the nonzero solution for the order parameter.
However, because of the superselection rules (resulting from the
conservation of the barionic charge) any $N$-particle system must
be in the Fock state -- the state with a well defined particle
number. Therefore the mean value of the boson field operator {\it
must vanish} in this state as the field operator changes the
number of particles.

In the following we use our model to demonstrate how to define the order
parameter, occupation of the condensate, and its fluctuations.
At zero temperature the system
is in the ground state and one might naively expect that it is totally Bose
condensated. However, the ground state of the N-particle bosonic system is not
equivalent to the Bose-Einstein condensate. Interactions can significantly
deplete the condensate. We are going to show this effect in the most
spectacular but also in relatively simple case of the zero temperature.

Let us now define the hierarchy of the reduced $s$-particle density matrices
which can be conventionally obtained by averaging the density matrix of the
total system of $N$ particles over the degrees of freedom of $N-s$ remaining
particles. For a given $N$-particle quantum state $\Psi({\bf X}_N)$ the
corresponding $s$-particle reduced density matrix $\rho_s({\bf X}_s; {\bf
Y}_s)$ is defined by:
\begin{equation}
\label{rho1}
\rho_s({\bf X}_s; {\bf Y}_s)=
\int {\rm d}{\bf R}_{N-s} \Psi^*({\bf X}_s,{\bf R}_{N-s})
\Psi({\bf Y}_s,{\bf R}_{N-s}).
\end{equation}
We use previously defined shorthand notation for vectors in a configuration
space of
$s$-particles. The reduced density matrix describes
the subsystem of $s$-particles and can be directly related to different
measurement processes. For the statistical description of the system
one should first of all define the statistical density matrix by averaging
all $N$-particle density matrices with the appropriate statistical weights
depending on the ensemble. In general it is quite a complicated task
but at zero temperature there is only one quantum state of the system and
no statistical averaging is necessary.

The total wave function (or density matrix) carries all the
information about the system. In real experiments however one does not probe
simultaneously all the particles. Typical detection scheme consists on the
measurement of one or at most few particles at a given time. In other words a
single measurement process is reduced to
a subsystem of small number of particles. Such subsystems are
described by reduced density matrices. In the considered here case of
zero temperature the description of the interacting system the $s$-particle
density matrix can be brought to the following form:
\begin{equation}
\label{s-part}
\rho_s({\bf X}_s; {\bf Y}_s) = \rho^{CM}({\bf x}^c_s, {\bf y}^c_s)
\Phi_0(\sqrt{\omega} {\bf X}^c_{s-1}) \Phi_0(\sqrt{\omega} {\bf Y}^c_{s-1}).
\end{equation}
The functions $\Phi_0$ describes the ground state of $s-1$
quasi-particles (collective  relative coordinates, see Eq.\
(\ref{phirel})) while $\rho^{CM}$ corresponds to the density
matrix of center of mass of the subsystem:
\begin{eqnarray}
\rho^{CM}({\bf x}^c_s, {\bf y}^c_s)&=&\left( \frac{{\omega}_s}{\pi}
\right)^{d/2}
{\rm exp}\left[
      \frac{\delta_s}{2} \, {\bf x}^c_s \,{\bf y}^c_s
         \right]  \nonumber \\
&&{\rm exp}\left[
-\frac{1}{2}
      \left(\omega_s + \frac{\delta_s}{2}\right)
      \left[({\bf x}^c_s)^2 + ({\bf y}^c_s)^2 \right]
\right]. \nonumber \\
&&
\end{eqnarray}
The $s$-particles collective coordinates are defined in the familiar way:
$({\bf X}^c_{s-1}, {\bf x}^c_s) = {\cal Q}_s \,{\bf X}_s$,
$({\bf Y}^c_{s-1},{\bf y}^c_s)= {\cal Q}_s \, {\bf Y}_s$ and frequencies
$\omega_s$, $\delta_s$ as well as auxiliary parameter $\gamma_s$ are:
\begin{eqnarray}
\gamma_s &=& 1 -\frac{s(1-\omega)}{N},\\
\omega_s &=& \frac{\omega}{\gamma_s}, \\
\delta_s &=& \left( \frac{1-\omega}{N} \right)^2 \frac{s(N-s)}{\gamma_s}.
\end{eqnarray}

Having defined the $s$-particle matrices we are ready now to
analyze the nature of the Bose-Einstein condensation of the
interacting system and to discuss the meaning of the order
parameter. To this end we write the density matrix Eq.\
(\ref{s-part}) in the diagonal form:
\begin{equation}
\label{spectral}
\rho_s({\bf X}_s; {\bf Y}_s) =
\sum_{\bf n} \lambda^{(s)}_{\bf n} \phi^{(s)}_{\bf n}({\bf X}_s)\,
\phi^{(s)}_{\bf n}({\bf Y}_s).
\end{equation}
The function $\phi^{(s)}_{\bf n}({\bf X}_s)$ can be treated as the wave
function of the $s$-particle subsystem:
\begin{equation}
\label{cm}
\phi^{(s)}_{\bf n}({\bf X}_s)=\Phi_0(\sqrt{\omega}{\bf X}^c_{s-1})
\Phi_{\bf n}(\sqrt{\alpha_s} {\bf x}^c_s),
\end{equation}
where $\Phi_0(\sqrt{\omega}{\bf X}^c_{s-1})$ is the ground state
wave function of the relative degrees of freedom. This function
corresponds to the ground state of $s-1$ noninteracting
quasi-particles (in $d$-spatial dimensions) subject to the
external harmonic potential of frequency $\omega$. The second part
of the Eq.\ (\ref{cm}) describes states of the center of mass of
$s$-particles; $\Phi_{\bf n}$ is simply the $d$-dimensional
harmonic oscillator wave function corresponding to the effective
center of mass frequency $\alpha_s$. Quantum numbers ${\bf
n}=(n_1, \ldots, n_d)$ label different states of the center of
mass while $n=n_1+ \ldots +n_d$ corresponds to the energy of the
given state. The effective center of mass frequency $\alpha_s$ is:
\begin{equation}
\alpha_s = \left[ \omega_s (\omega_s + \delta_s) \right]^{1/2}.
\end{equation}
It is interesting to observe that all the frequencies of the
relative motion of the $s$-particles subsystem are exactly the
same as the frequencies of the relative motion of the whole
system, i.e., equal to $\omega$. On the other hand the center of
mass oscillation frequency of the subsystem is neither equal to
$\omega$ nor to 1 (trap frequency). This collective degree of
freedom couples to the center of mass of $N-s$ remaining particles
what leads to some energy shift. Finally, the eigenvalues
$\lambda^{(s)}_{\bf n}$ of $\rho_s$ are equal to the occupation
probabilities of a given $s$-particle state:
\begin{equation}
\label{lambda} \lambda^{(s)}_{\bf
n}=\left(\frac{\omega_s}{\alpha_s}\right)^{d/2} \left(\frac{2
\sqrt{\omega_s \alpha_s}}{\omega_s+\alpha_s}\right)^d
\left(\frac{\alpha_s-\omega_s}{\alpha_s+\omega_s}\right)^n.
\end{equation}
It follows from the normalization condition for the density matrix
that $\sum_{\bf n} \lambda^{(s)}_{\bf n} =1$.

The spectral decomposition of the reduced single particle density matrix gives
natural single-particles states $\phi^{(1)}_{\bf n}({\bf x})$. These
states are crucial for the definition of the condensate wave function (order
parameter). It can be seen from Eq.(\ref{lambda}) that if $N$ goes to
infinity (thermodynamic limit) with fixed value of the interaction frequency
$\omega$ the lowest eigenvalue $\lambda^{(1)}_0$ dominates the others:
\begin{eqnarray}
\lambda^{(1)}_0       &\simeq& 1,\\
\lambda^{(1)}_{\bf n} &\simeq& \left(\frac{(1-\omega)^2}{4N}\right)^n,
\phantom{123}{\rm if}\phantom{123} n \neq 0.
\end{eqnarray}
This behavior signifies nothing else but the onset of the Bose-Einstein
condensation. The single particle density matrix becomes very close to the pure
state because with quite good accuracy it can be approximated by $\rho_1( {\bf
x}, {\bf y}) \approx \phi^{(1)}_0({\bf x})\, \phi^{(1)}_0({\bf y})$. This
particular single-particle ground state $\phi^{(1)}_0({\bf x})$ is usually
called the order parameter. The $N$-particle wave function can be quite
accurately approximated by the $N$-fold product of the order parameter.

Above the critical temperature the situation is completely
different, namely all eigenvalues of $\rho_1$ should be close to zero
what means that the single particle reduced density matrix is `very far' from
the pure state and order parameter vanishes -- there is no leading state in the
spectral decomposition of the single-particle density matrix.
\begin{figure}
   \begin{center}
   \epsfxsize 7.5cm
   \epsffile{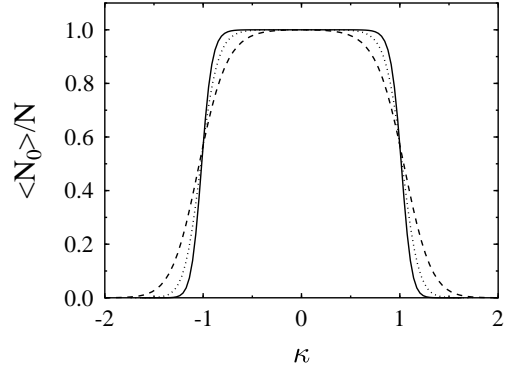}
   \end{center}
   \caption{
            Mean occupation of the condensate plotted as a function
            of the parameter $\kappa = \log\omega/\log N$ for different
            number of
            particles; $N=10^3$ -- dashed line; $N=10^5$ -- dotted line,
            and $N=10^8$ -- full line.
            }
   \label{fig1}
\end{figure}
Our analytic formula allows to study quantitatively the role of interactions on
the Bose-Einstein condensate.  On the basis of the discussion it is obvious
that the average occupation of the condensate becomes:
\begin{equation}
\label{n0}
\langle N_0 \rangle = N \int {\rm d}{\bf x} {\rm d}{\bf y}
\phi^{(1)}_0 ({\bf x})  \rho_1({\bf x}, {\bf y})
{\phi^{(1)}_0}({\bf y}) = N \lambda^{(1)}_{0}.
\end{equation}
In the case of the ideal gas at zero temperature the above equation gives, of
course, $\langle N_0 \rangle = N$; all particles occupy the single particle
ground state. For a fixed number of particles, if the interaction strength
$|\log\omega|$ grows, the occupation of the condensate decreases. This
behavior is presented in Fig. 1 where we show the mean occupation of the
condensate versus the exponent $\kappa= \log\omega/ \log N$ for different
values of particle number $N$ in three spatial dimensions ($d=3$). The values
of $\kappa$ less than zero signify repulsive interactions while $\kappa > 0$
corresponds to attraction. One can easily see that if the interaction becomes
strong ($|\kappa| \simeq 1$) the condensate is almost totally depleted. All
curves presented in the figure tend to an universal curve if the number
of particles increases.  When $N$ increases to infinity with $\kappa$ being
constant then our expression for the occupation of the condensate has the form:
\begin{equation}
\frac{\langle N_0 \rangle}{N}=
\left(\frac{2}{1+\sqrt{N^{\kappa-1}+N^{-(\kappa+1)}+1}}\right)^d.
\end{equation}
The above formula, valid in the thermodynamic limit,  gives an universal
critical behavior. It exhibits no depletion ($N_0 = N$) for
$|\kappa| <1$ followed  by an abrupt jump and total destruction of the
condensate  ($N_0=0$)  for $|\kappa|>1$.

The effect of quantum depletion of the trapped atomic condensate
with a short range interactions, for the realistic experimental
parameters, has been estimated to be of the order of $1\%$
\cite{Dalfovo}. This is opposite to the case of superfluid helium
where this effect accounts for depletion as large as more  than
$90\%$ \cite{Huang}. Our model exhibits very interesting feature.
It shows that in large $N$ limit the quantum effects are almost
negligible or totally destroy the condensate depending on the
value of the interaction strength. At this point it is not clear
if this is an unique feature of our model or if it is a more
general result.

We see that interactions play an important role. If they are
strong, the condensate disappears although the $N$-particle system
remains in its ground state. There is no coherence in the strongly
interacting system, i.e., no wave function can be assign to a
single particle subsystem. At this point we want to make a comment
about the notion of the coherence of the Bose-Einstein condensate.
Approximate methods assume explicitly that the mean value of the
boson field operator is different from zero in the case of the
Bose-Einstein condensate. Therefore, the folk wisdom associates
the condensate with the coherent state -- the analog of the
coherent state of the electromagnetic field. This analogy is of
limited value and in fact may be misleading because the condensate
must be in a Fock state in which a mean value the field operator
vanishes. However there is coherence in the condensate in the
sense that majority of particles are described by the same wave
function with the same phase. The expression for this wave
function can be obtained rigorously only when one considers the
single particle reduced density matrix.

The 2-particle reduced density matrix allows to find a joint probability of
finding one particle in a given single particle state and simultaneously
another particle in another given state. In particular we have:
\begin{eqnarray}
\langle N_0 (N_0 -1) \rangle &=& N(N-1)
\int {\rm d}{\bf X}_2 {\rm d}{\bf Y}_2\,
\rho_2({\bf X}_2, {\bf Y}_2) \nonumber \\
&&\phi^{(1)}_{0}({\bf x}_1)  \phi^{(1)}_{0}({\bf x}_2)
{\phi^{(1)}_{0}}({\bf y}_2) {\phi^{(1)}_{0}}({\bf y}_1).
\end{eqnarray}
Simple integration gives:
\begin{eqnarray}
\langle N_0 (N_0 -1) \rangle &=&N(N-1)
\left( \frac{2 \sqrt{\omega \alpha_1}}{\omega + \alpha_1}\right)^d
\left( \frac{2 \sqrt{\omega_2 \alpha_1}}{\omega_2 + \alpha_1} \right)^d
\nonumber \\
&&\times \left(
\frac{\omega_2 +\alpha_1}{\omega_2+\alpha_1+\delta_2}
\right)^{d/2}.
\end{eqnarray}
Now we are ready to analyze the fluctuations of the condensate defined as:
\begin{equation}
\label{fluk} \langle \delta^2 N_0 \rangle = \langle N_0^2 \rangle
- \langle N_0 \rangle^2.
\end{equation}
These fluctuations are shown in Fig.2. We see that as the
interaction strength grows up (at fixed number of particles) the
fluctuations start to grow from zero value for the ideal gas.
However when the interactions become so strong that condensate
practically disappears ($|\kappa| \simeq 1$) fluctuations also
decrease - as there is no condensate the fluctuations also die
out. The fluctuations are maximal in a region of the critical
destruction of the condensate by quantum effects.
\begin{figure}
   \begin{center}
   \epsfxsize 7.5cm
   \epsffile{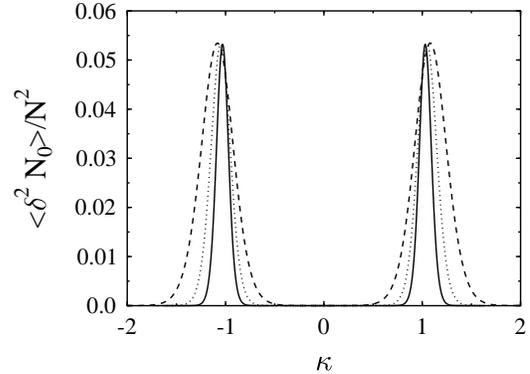}
   \end{center}
   \caption{
            Fluctuations of the condensate plotted as a function
            of the parameter $\kappa = \log\omega/\log N$ for different
            number of
            particles; $N=10^3$ -- dashed line; $N=10^5$ -- dotted line,
            and $N=10^8$ -- full line.
            }
   \label{fig2}
\end{figure}

\subsection{Finite temperatures}
In this subsection we estimate, within the exact model, some
effects in finite-temperature behavior of the trapped gas.
Rigorous description of the condensate requires a knowledge of the
statistical density matrix of the $N$-particle system. Knowing
this matrix one can apply the procedure described previously to
define the finite temperature condensate, its occupation and
fluctuations. However, because of a huge degeneracy of high energy
states the statistical averaging procedure is difficult.
Therefore, we limit our study to the case of weak interactions
($|\kappa| < 1$) when we can neglect the quantum effects leading
to a significant depletion of the condensate. In this case we can
expect that the condensate wave function in a finite temperature
is equal to the ground state of the harmonic oscillator with some
effective frequency (characterizing a mean field experienced by a
single particle) which at $T=0$ is equal to $\alpha_1$. In the
case of a weak interaction this frequency can be approximated by
$\alpha_1 \approx \omega$. We expect that, similarly to the
noninteracting case, the main effect of the temperature is to
deplete the condensate rather than modify the condensate wave
function, i.e., the frequency $\alpha_1$.

As it has been shown in \cite{Magda}, the trace of the density
matrix of the $N$-particle system, i.e., the microcanonical
partition function $\Gamma(N,E)$ is identical (in the
thermodynamic limit) to the microcanonical partition function
$\Gamma_0(N,E,\omega)$ of the $N$ noninteracting bosonic particles
(quasi particles) trapped by the harmonic potential of frequency
$\omega$:
\begin{equation}
\Gamma(N,E) \approx \Gamma_0(N,E,\omega).
\end{equation}
The effect of the center-of-mass excitations on the spectrum and
on the statistical properties of the system is negligible since it
is related to only one degree of freedom as compared to the $N-1$
remaining collective degrees of freedom.  This fact, together with
our remarks about the condensate wave function, signifies that the
system of interacting (via harmonic forces) particles is, in the
thermodynamic limit, equivalent to the ideal gas in the
oscillatory trap.

This observation allows to recall all results obtained for the ideal Bose gas
\cite{Bagnato}. In particular, considered here interacting system undergoes the
Bose-Einstein condensation at the temperature $T_c$ equal to:
\begin{equation}  \label{Tc}
T_c = \omega \left( \frac{N}{\zeta(3)} \right)^{1/3},
\end{equation}
where the $\zeta$ is the Riemann function, $\zeta(3)=1.2020569$. This critical
temperature should be compared to the critical temperature of the
noninteracting system,
\begin{equation}
T_0 = \left( \frac{N}{\zeta(3)} \right)^{1/3}.
\end{equation}
The shift of the critical temperature $\Delta T = T_c-T_0$ is therefore equal
to:
\begin{equation}
\frac{\Delta T}{T_0} = \omega -1.
\end{equation}
It is negative in the case of repulsive interaction and has the
opposite sign for the attractive system what is in a qualitative
agreement with results of Ref.\ \cite{Stringari} for the system of
trapped atoms. The critical temperature for the trapped gas with
repulsive interactions is decreased as compared to the
noninteracting case: due to  interactions a mean separation
between particles grows, therefore the quantum statistical effects
become important at larger de Broglie wavelength.

It is worth to stress that the transition temperature for the system with short
range interactions remains a controversial subject even in the uniform case
\cite{Huang_t}.  There is no consensus how the shift of the temperature should
depend on the interaction strength, nor even the sign.

The fraction of condensate particles $\langle N_0\rangle$ is:
\begin{equation}
\langle N_0\rangle = N - \left( \frac{T}{\omega} \right)^3 \zeta(3).
\end{equation}
It is known that the critical temperature and mean occupation of the ground
state do not depend on the statistical ensemble used for the description of
the system. The value of the fluctuations of the number of the particles in
the condensate calculated for different statistical ensembles differ
significantly \cite{Gajda}. In the canonical ensemble the square of the
fluctuations
$\langle\delta^2 N_0\rangle _{{\rm CN}}= \langle N_0^2\rangle_{\rm CN} -
\langle N_0\rangle ^2_{{\rm CN}}$ is:
\begin{equation}  \label{CN}
\langle \delta^2 N_0\rangle_{{\rm CN}} = \left( \frac{T}{\omega} \right)^3
\zeta(2),
\end{equation}
where $\zeta(2)=\pi^2/6$.
In the case of perfectly isolated system
(microcanonical ensemble) fluctuations of the ground state occupation are
smaller than canonical ones \cite{Gajda}:
\begin{equation}  \label{MC}
\langle \delta^2 N_0 \rangle_{{\rm MC}} = \left( \frac{T}{\omega} \right)^3
\zeta(2) \left( 1 - \frac{3 \zeta^2(3)}{4 \zeta(4) \zeta(2)} \right) ,
\end{equation}
and $\zeta(4)=\pi^4/90$.

\section{Bogoliubov approximation}

\subsection{Ground state}
In realistic cases the exact solution of the many-body Schr{\"o}dinger
equation is impossible. Instead, mean-field approaches are being developed. The
basic idea of the mean-field theory was introduced by Bogoliubov.
We will now formulate the Bogoliubov approximation in the case of
inhomogeneous systems \cite{Esry}. Next we find the energy spectrum within this
approximation for our exactly solvable model. In order to do this we rewrite
the Hamiltonian (\ref{a0}) using the second quantization formalism. Thus we
introduce the field operator $\hat{\psi}({\bf x})$ which annihilates a
particle at a point ${\bf x}$ and its conjugate $\hat{\psi}^{\dag }({\bf x})$
which creates a particle at a point ${\bf x}$. These operators fulfill
standard bosonic commutation relations:
\begin{equation}
\left[ \hat{\psi}({\bf x}),\hat{\psi}^{\dag }({\bf x}^{\prime })\right]
=\delta ({\bf x}-{\bf x}^{\prime }).
\end{equation}
In the second quantization formalism the Hamiltonian becomes:
\begin{eqnarray}
\hat{H} & = &\int {\rm d}{\bf x} \hat{\psi}^{\dag}({\bf x})
H_0({\bf x}) \hat{\psi}({\bf x}) \nonumber \\
&&+ \frac{1}{2}\int {\rm d}{\bf x}  d{\bf x}^{\prime}\,
\hat{\psi}^{\dag}({\bf x}) \hat{\psi}^{\dag}
({\bf x}^{\prime}) V({\bf x}-{\bf x}^{\prime})
\hat{\psi}  ({\bf x}) \hat{\psi} ({\bf x}^{\prime }),
\label{a1}
\end{eqnarray}
where
\begin{equation}
H_0({\bf x)}=\frac 12({\bf p}^2+{\bf x}^2).  \label{a2}
\end{equation}

The Bogoliubov approximation is formulated in two steps. The first one
is to express the field operator as a sum of its mean value $\sqrt{N_0}\phi
_0({\bf {x})}$ and an operator $\hat{\phi}({\bf x})$ responsible for the
fluctuations around the mean value. Below the Bose-Einstein condensation
temperature the occupation of the ground state is nonzero. It is
therefore convenient to write
\begin{equation}
\hat{\psi}({\bf x})=\sqrt{N_0}\phi _0({\bf x})+\widehat{\phi }({\bf x}).
\label{a5}
\end{equation}
Note that now the mean value of the field operator $\hat{\psi}({\bf x})$ is not
equal to zero and is proportional to the square root of the number of condensed
particles $N_0$. The spirit of the Bogoliubov approximation is based on the
fact that the occupation of the condensate is of the order of total particles
number $N_0\simeq {\cal O}(N)$. This, in principle, limits the Bogoliubov
approach to low temperatures. In the following, consistently with the above
assumption we will substitute in the Eq.(\ref{a5}) $N_0$ by $N$.  For any
physically realizable $N$-particle state, and in particular for our solutions
of the system Eq.(\ref{a0}), the mean value of the field operator is zero.
(Strictly, separation of the Eq.(\ref{a5}) should be done for operators
conserving particle number. However, extracting a $c$--number part of the field
operator has formally the same consequences and is easier to handle).  If such
a form is substituted into the total Hamiltonian Eq.(\ref{a1}) the number of
particles is no longer conserved. To overcome this difficulty one considers the
grand canonical Hamiltonian instead
\begin{equation}
\widehat{K}=\widehat{H}-\mu \widehat{N},  \label{hamiltonian}
\end{equation}
where $\widehat{N}$ is the total particle number operator and $\mu$ is the
chemical potential. It should be chosen in such a way that the
mean particle number is equal to the desired value. Self consistent equation
for the condensate wave function $\phi _0({\bf x})$ follows from the
assumption that the decomposition (\ref{a5}) gives the best self consistent
function. We find:
\begin{equation}
\label{a71}
\left\{ H_0({\bf x}) + V_{eff}[\phi _0,{\bf x}] \right\} \phi _0({\bf x})
=\mu \phi _0({\bf x}),
\end{equation}
where the effective potential is:
\begin{equation}
V_{eff}[\phi _0,{\bf x}]= \frac{\sigma}{2} N \Omega^2 \int {\rm d}
{\bf x}^{\prime} \phi_0^{*}({\bf x}^{\prime })({\bf x}-{\bf
x}^{\prime })^2 \phi _0({\bf x}^{\prime }).
\end{equation}
This equation replaces the standard Gross-Pitaevskii equation for the
condensate wave function. The effective potential in the Eq.(\ref{a71}) has
different form than usuall nonlinear term appearing in the Gross-Pitaevskii
equation because of long  range forces assumed in our model as opposed to
the more realistic zero range interactions.

The lowest energy state of the Hamiltonian from Eq.(\ref{a71}) is
\begin{equation}
\label{Bwf}
\phi _0({\bf x})=\left( \frac \omega \pi \right) ^{\frac 34}\exp
(-\frac{1}{2} \omega {\bf x}^2),
\end{equation}
and the value of chemical potential is $\mu =3/4(1+\omega)$.
The function Eq.(\ref{Bwf}) is the Bogoliubov approximation to the exact order
parameter $\phi^{(1)}_0(\sqrt{\omega}{\bf x})$ found in the previous section:
\begin{equation}
\phi^{(1)}_0(\sqrt{\omega}{\bf x}) =\left( \frac{\alpha_1}{\pi} \right)^{\frac
34}\exp (-\frac{1}{2} \alpha_1 {\bf x}^2),
\end{equation}
where the effective frequency $\alpha_1$ in the limit of weak interactions
can be approximated by:
\begin{equation}
\alpha_1 \approx \omega \left(1 + \frac{1-\omega}{N} \right).
\end{equation}
Because the effective frequency $\alpha_1$ is very close to $\omega$ the
Bogoliubov expression for the condensate wave function is quite accurate in the
limit of weak interactions. However, if the interaction strength is large the
Bogoliubov method fails to reproduce the condensate wave function. This is
consistent with the basic assumption of the Bogoliubov method which requires
the order parameter (multiplied by the mean occupation of the condensate)
to be large. As we have shown in the previous
section this is not the case for strongly interacting system.

\subsection{Excitation spectra}
The second step in the Bogoliubov method is to find the low energy
excitation spectrum of the system by expanding the total Hamiltonian around
the mean value of the field operator given by the solution Eq.(\ref{Bwf}).
After substituting the field operator Eq.(\ref{a5}) into $\widehat{K}$ and
retaining all terms up to ${\cal O}(\widehat{\phi }^2)$, the operator
$\widehat{K}$ can be diagonalized with the help of a canonical transformation:
\begin{equation}
\widehat{\phi }({\bf x})=\sum_\lambda \left( u_\lambda ({\bf x})\beta
_\lambda +v_\lambda ^{*}({\bf x})\beta _\lambda ^{\dag }\right) .  \label{a6}
\end{equation}
where $\beta _\lambda$ and $\beta _\lambda ^{\dagger }$ are bosonic
annihilation and creation operators. The diagonal form of the operator
$\widehat{K}$ is
\begin{equation}
\widehat{K}=\sum_\lambda \Delta _\lambda \int {\rm d}{\bf x}
v_\lambda ^{*}({\bf x})v_\lambda ({\bf x})+\sum_\lambda \Delta
_\lambda \beta _\lambda ^{\dagger }\beta _\lambda ,
\end{equation}
provided that functions $U_\lambda ({\bf x})$ and $V_\lambda ({\bf x})$ defined
as:
\begin{eqnarray}
U_\lambda ({\bf x}) &=&u_\lambda ({\bf x})+v_\lambda ({\bf x}), \\
V_\lambda ({\bf x}) &=&u_\lambda ({\bf x})-v_\lambda ({\bf x}),
\end{eqnarray}
satisfy the normal--mode equations:
\begin{eqnarray}
H_\omega ({\bf x}) U_\lambda ({\bf x})
&+&\int {\rm d}{\bf x}^{\prime }G({\bf x},{\bf x}^{\prime })U_\lambda ({\bf x}%
^{\prime })=\Delta _\lambda V_\lambda ({\bf x}),  \label{bog1} \\
H_\omega ({\bf x})  V_\lambda ({\bf x})
&=&\Delta _\lambda U_\lambda ({\bf x}) \label{bog2},
\end{eqnarray}
with the following normalization condition:
\begin{equation}
\int {\rm d}{\bf x}\left( u_\lambda ^{\star }({\bf x})u_{{\lambda
}^{\prime
}}({\bf x})-v_\lambda ^{\star }({\bf x})v_{{\lambda }^{\prime }}({\bf x})\right) =\delta _{{%
\lambda },{\lambda }^{\prime }}.
\end{equation}
In these formulas $\Delta _\lambda $ has a meaning of an eigenvalue,
$H_\omega ({\bf x})=1/2({\bf p}^2+\omega ^2{\bf x}^2)-3\omega/2$, and the
integral kernel $G({\bf x},{\bf x}^{\prime })$ is:
\begin{equation}
G({\bf x},{\bf x}^{\prime })=2N\phi _0({\bf x})V({\bf x}-{\bf x}^{\prime
})\phi _0({\bf x}^{\prime }).
\end{equation}
The above equations can be easily solved if we expand functions
$U_\lambda ({\bf x})$ and $V_\lambda ({\bf x})$ in the basis of
eigenfunctions $\psi _{\bf n}({\bf x})=\psi _{n_xn_yn_z}({\bf x})$
of the Hamiltonian $H_\omega ({\bf x})$:
\begin{eqnarray}
U_\lambda ({\bf x}) &=&\sum_{{\bf n}}a_{{\bf n}}^\lambda
\psi _{\bf n}(\sqrt{\omega}{\bf x}),\\
V_\lambda ({\bf x}) &=&\sum_{{\bf n}}b_{{\bf n}}^\lambda
\psi _{{\bf n}}(\sqrt{\omega}{\bf x}),  \label{a11b}
\end{eqnarray}
where components of the vector ${\bf n}=(n_x,n_y,n_z)$ are the standard
quantum numbers of the harmonic oscillator eigenfunction of energy equal to
$n\omega $, where $n=n_x+n_y+n_z$.\newline

The coefficients $a_{{\bf n}}^\lambda $ and $b_{{\bf n}}^\lambda $
have to be determined from the Eqs.\ (\ref{bog1}) and
(\ref{bog2}). Let us remind that the solution of the Eq.\
(\ref{a71}) for the order parameter $\phi _0({\bf x})$ is the
first function of the chosen basis set, $\phi _0({\bf x})=
\psi _{000}(\sqrt{\omega}{\bf x})$. For this reason and also due to the
oscillatory form of the inter-particle interactions, the integral
kernel in Eqs.\ (\ref{bog1}) and (\ref{bog2}) couples only these
basis functions $\psi _{{\bf n}}$ which correspond to three lowest
eigenstates ($n=0,1$ and 2) of the Hamiltonian $H_\omega ({\bf
x})$. For larger $n$ quasi-particles excitation energies are those
of the harmonic spectrum:
\begin{equation}
\Delta _n=n\omega ,{\rm \phantom{123}if\phantom{12}}n>2.
\end{equation}
These eigenvalues are
degenerated. In general, there is a close link between the number of
eigenmodes $u_\lambda ({\bf x})$ and $v_\lambda ({\bf x})$ corresponding to
the eigenvalue $n\omega $ and the number of the oscillatory states of the
same energy. Therefore, in order to classify independent solutions, it is
convenient to use oscillatory quantum numbers instead the parameter $\lambda$
which simply enumerates quasi-particles eigenmodes. With this notational
modification the solutions of the Eq.(\ref{bog1},\ref{bog2}) corresponding to
energies  $\Delta _n$ with $n>2$ are the following:
\begin{eqnarray}
u_{{\bf n}}({\bf x}) &=&\psi _{{\bf n}}(\sqrt{\omega}{\bf x}), \\
v_{{\bf n}}({\bf x}) &=&0.
\end{eqnarray}
The low laying interacting states involve coupling of the bare oscillatory
eigenfunctions. For the lowest excitation energy we get:
\begin{equation}
\Delta _1=1.
\end{equation}
There are three different modes of that energy corresponding to
the excitation of one of the $x$, $y$, or $z$ degree of freedom.
Below we present only one pair of the eigenmodes as the remaining
two can be obtained by the permutation of the indices only. The
$x$-direction eigenmodes are:
\begin{eqnarray}
u_{100}({\bf x}) &=&\frac{(1+\omega )}{\sqrt{4\omega }}
\psi _{100}(\sqrt{\omega}{\bf x}),\\
v_{100}({\bf x}) &=&\frac{(1-\omega )}{\sqrt{4\omega }}
\psi _{100}(\sqrt{\omega}{\bf x}).
\end{eqnarray}
Let us notice that energy of this mode of excitations is equal to the single
excitation quantum of the trap mode.

The second excitation energy is equal to one of the excitation energy
of the Hamiltonian $H_{\omega}({\bf x})$, namely:
\begin{equation}
\Delta_2 = 2 \omega.
\end{equation}
There are six different eigenmodes of the above energy which is exactly the
degeneracy of the second state of the 3D oscillator. The first three pairs
of them are related to the double excitation along one of the axis of the
coordinate system and are of the form:
\begin{eqnarray}
u_{200}({\bf x})&=& \psi_{200}(\sqrt{\omega}{\bf x}) +
\frac{\omega^2 -1}{2 \sqrt{2}
\omega} \psi_{000}(\sqrt{\omega}{\bf x}), \\
v_{200}({\bf x})&=&-\frac{\omega^2-1}{2 \sqrt{2} \omega}
\psi_{000}(\sqrt{\omega}{\bf x}),
\end{eqnarray}
(the other two pairs of eigenstates can obtained by the permutation of the
oscillatory quantum numbers, as previously). The remaining three pairs of
eigenmodes correspond to two single quanta of excitations along two
different principal axis of the coordinate system.
For example, one such pair is:
\begin{eqnarray}
u_{110}({\bf x}) &=& \psi_{110}(\sqrt{\omega}{\bf x}), \\
v_{110}({\bf x}) &=& 0,
\end{eqnarray}
and two others can be obtained by permutations of
indices. 

The shift in the ground state energy is given by
\begin{equation}
\sum_{\bf n} \, \Delta _{n}\int {\rm d} x\,v_{\lambda }^2(x)=\frac{(1-\omega
)^2}{4\omega }(2+2\omega +\omega ^2).  \label{shift}
\end{equation}

The excitation energies $\Delta _n$ obtained within the Bogoliubov approach
are the same as exact eigenenergies of the interacting system. The
degeneracies of the eigenenergy state when we compare with Ref.\cite{Magda}
are also the same. The Bogoliubov method is very well suited for the
description of the excitation spectrum of the quantum degenerate gas. This
result is somewhat surprising. One might rather expect that Bogoliubov
approach works well only for the short range interaction. Our calculation
shows that it works also in a rather exotic case when the interaction strength
grows quadratically with the distance between particles. There are some
differences between wave functions obtained in the Bogoliubov approximation
and exact solutions of the $N$-particle Hamiltonian \cite{Magda}. They come
from the fact that the exact solution cannot be written as a symmetrized
product of any single particle functions.

\subsection{Condensate fraction and fluctuations}

In this subsection we consider Bose gas at finite temperatures. We
will study the impact of interactions on the occupation of the
condensate and its fluctuations using Bogoliubov method
\cite{Giorgini}. We will compare the obtained results with those inferred from the exactly soluble model.

The statistical density matrix is $\rho =Z^{-1}\exp
(-\widehat{K}/T)$, where $\widehat{K}$ is the Hamiltonian given in
Eq.(\ref{hamiltonian}) and $Z$ is the statistical sum. This
density matrix describes the excited subsystem only
(quasi-particles). Number of quasi-particles is not conserved and
the condensate is assumed to act as a reservoir of
quasi-particles. In fact all this assumptions are in the spirit of
the Maxwell's demon ensemble introduced for the description of the
ideal gas below the condensation temperature \cite{Gajda}.
Imposing the constraint on the total number of particles
implies that occupation and fluctuations of the condensate can be
directly related to the mean number and fluctuations of
quasi-particles.

The mean number of excitations (quasi particles) above the condensed phase
$\langle N_e \rangle = N-\langle N_0 \rangle $ is defined as:
\begin{equation}
\langle N_e \rangle=\int {\rm d}{\bf x}\, \langle\widehat{\varphi
}^{\dagger}({\bf x}) \widehat{\varphi}({\bf x})\rangle,
\end{equation}
what leads to the following expression:
\begin{equation}
\label{mean} \langle N_e \rangle =\sum_{{\bf n}\neq0} \int {\rm
d}{\bf x} \left\{ \left[ u_{\bf n} ^2({\bf x})+v_{\bf n} ^2({\bf
x})\right] f_{\bf n} +v_{\bf n} ^2({\bf x})\right\},
\end{equation}
where $f_{\bf n} =[\exp(\Delta_{\bf n} /T)-1]^{-1}$ plays the role
of the mean quasi-particle occupation of the given energy state.
The functions $u_{\bf n}$ and $v_{\bf n}$ are closely related to
the wave functions of harmonic oscillator. The integration can be
easily performed but the summation over all eigenstates might be
quite difficult because of huge degeneracy of the energy levels.
Fortunately, in our case almost all functions $v_{\bf n}$ vanish
and their contribution to the final result is negligible at finite
temperatures. Therefore the problem can be easily reduced to the
calculation of the canonical occupation of the condensate trapped
in the harmonic trap of frequency $\omega$. Again, similarly as in
the exact solution we will treat separately the two regimes: (i)
zero temperature limit where only quantum effects described by the
last term of the Eq.\ (\ref{mean}) affect the condensate
population, (ii) finite temperature case where the above mentioned
effect is negligible. The occupation of the interacting condensate
at zero temperature is:
\begin{equation}
\label{qdep}
\langle N_0 \rangle = N -\frac 3{4\omega }(1-\omega)^2 -\frac
3{8\omega^2}(1-\omega^2)^2.
\end{equation}
The first term in Eq.\ (\ref{qdep}) corresponds to the weak
interaction limit of the exact result, which is
\begin{equation}
\langle N_0 \rangle  \approx N - \frac{3}{4\omega}(1-\omega)^2.
\end{equation}
The second term in Eq.\ (\ref{qdep}) is new. This additional term
overestimates the depletion of the condensate as compared to the
rigorous treatment. Therefore the Bogoliubov treatment does not
describe quantitatively an occupation of the zero temperature
condensate.

On the other hand, in the thermodynamic limit, we recover
the familiar expression:
\begin{equation}
\label{n01}
\langle N_0 \rangle = N - \left( \frac{T}{\omega} \right) ^3\zeta (3).
\end{equation}
This is exactly the same result which we obtained in the exact
treatment. Similarly, the critical temperature which can be
defined by setting $\langle N_0 \rangle $ to zero in the Eq.\
(\ref{n01}):
\begin{equation}
T_c = \omega \left(\frac{N}{\zeta(3)}\right)^{1/3},
\end{equation}
is identical to the exact result, Eq.\ (\ref{Tc}). It might be
somewhat surprising but the Bogoliubov method works in our case
pretty well up to the critical temperature, i.e., in the region
where, in principle, its assumptions are not valid.

In a similar way we can obtain the fluctuations of the condensate
population. Because fluctuations, contrary to the mean occupation,
depend on the statistical ensemble, we use a notation which
explicitly indicates the kind of performed averages (i.e., the
canonical one). As we imposed constraint on the total number of
particles, these fluctuations are equal to the fluctuations of the
above-condensate part, $\langle \delta N_0^2\rangle_{\rm CN}
=\langle N_0^2 \rangle_{\rm CN} - \langle N_0 \rangle_{\rm CN}^2$:
\begin{eqnarray}
\langle \delta N_0^2 \rangle_{\rm CN} &=& \sum_{{\bf n}\neq 0}
\left\{ \int {\rm d}{\bf x}  \left[ u_{\bf n}^2({\bf x})+v_{\bf
n}^2({\bf x}) \right]^2 (f^2_{\bf n} + f^{}_{\bf n}) \right.
\nonumber
\\ &&\left.+ 4\left[ \int {\rm d}{\bf x} u_{\bf n}^{}({\bf
x})v_{\bf n}^{}({\bf x})
         \right]^2
\big( f^2_{\bf n} + f^{}_{\bf n}+1 \big)
\right\},
\end{eqnarray}
where the last term is responsible for the quantum fluctuations which do not
vanish at zero temperature.  Again the problem of calculating the condensate
fluctuation can be reduced to finding the fluctuations of the ideal trapped
Bose gas. In the case of zero temperature we have:
\begin{equation}
\langle \delta N_0^2 \rangle_{\rm CN} =
\frac{3}{2\omega^2}(1-\omega ^2)^2+ \frac 3{(4\omega ^2)^2} \left(
1-\omega ^2\right)^4.
\end{equation}
This formula overestimates (quite significantly) the condensate
fluctuations due to the quantum effects as compared to the exact
results in the limit of weak interaction, Eq.\ (\ref{fluk}).

The fluctuations  (in the thermodynamic limit and after neglecting the small
quantum fluctuations obtained above) at finite temperatures are:
\begin{equation}
\label{fluct} \langle \delta N_0^2 \rangle_{\rm CN}=
\left(\frac{T^{}}\omega \right) ^3\zeta (2).
\end{equation}
This result gives the correct value of the thermal fluctuations of
the condensate. The Bogoliubov method works very well in
predicting the thermal fluctuations of the condensate while it
fails to reproduce the correct value of the quantum fluctuation.
It might be surprising as in fact the approach should work in the
low temperature region. In our opinion, it is the substitution of
the operator annihilating the lowest energy state by a $c$-number
that is responsible for inaccurate treatment of some quantum
effects, particularly important at zero temperature.

\section{Conclusions}
In our paper we used the exactly soluble many-particle model to
illustrate the rigorous procedure of defining  the condensate phase
at zero temperature. By diagonalizing one-particle reduced density
matrix we were able to study in details the role of interactions
on the condensate at zero temperature. If the interaction strength
becomes large $|\kappa| >1$ condensate disappears even when the
system is in its ground state.  This total depletion of the
condensate has, in the thermodynamic limit, a character of
critical phenomena. The destruction of the condensate signifies a
breakdown of the mean-field theory. Due to strong quantum
correlations the system cannot be viewed as being composed of
independent quasi-particles moving in some effective potential
resulting from interactions with a rest of the system. Instead the
quasi-particles become strongly correlated and simple
single-particle picture is not longer valid.

We have also carefully compared the exact quantum solutions of the
oscillatory model with the approximate solutions obtained with the
help of the Gross-Pitaevskii equation and Bogoliubov
approximation. We have found that many of the characteristics of
the exact solutions like the excitation spectrum, occupation of
the condensate, and its thermal fluctuations are indeed reproduced
with the help of the approximate methods. The Bogoliubov approach
fails in the case of very strong interactions when the condensate
is almost destroyed.  This is however consistent with the basic
assumption of the Bogoliubov method which explicitly assumes small
condensate depletion and relays on the validity of a mean-field
description.  Surprisingly, in the studied model, the method works
quite well even at temperatures close to the critical one. On the
other hand the zero temperature (quantum) depletion and
fluctuations of the condensate are not given correctly by the
Bogoliubov method. Substitution of the destruction operator of the
particle in the lowest energy state by a classical field results
in inaccurate description of some quantum effects. Fortunately for
the weakly interacting system, these effects are small and can be
neglected.

\acknowledgments This work was supported by the KBN grant 2 P03B
130 15.

%\begin{figure}
%   \begin{center}
%   \epsfxsize 7.5cm
%   \epsffile{figure1.eps}
%   \end{center}
%   \caption{
%            Mean occupation of the condensate plotted as a function
%            of the parameter $\kappa = \log\omega/\logN$ for different
%            number of
%            particles; $N=10^3$ -- dashed line; $N=10^5$ -- dotted line,
%            and $N=10^8$ -- full line.
%            }
%   \label{fig1}
%\end{figure}
%
%\begin{figure}
%   \begin{center}
%   \epsfxsize 7.5cm
%   \epsffile{figure2.eps}
%   \end{center}
%   \caption{
%            Fluctuations of the condensate plotted as a function
%            of the parameter $\kappa = \log\omega/\log N$ for different
%            number of
%            particles; $N=10^3$ -- dashed line; $N=10^5$ -- dotted line,
%            and $N=10^8$ -- full line.
%            }
%   \label{fig2}
%\end{figure}

\end{document}